\documentclass[amsmath,amssymb,aps,prb,twocolumn,superscriptaddress]{revtex4}
\usepackage{bm}
\usepackage{epsfig}
\usepackage[usenames]{color}

\newcommand{\el}{{\rm el}}
\newcommand{\inel}{{\rm inel}}
\newcommand{\nm}{{\rm nm}}
\newcommand{\sgn}{\operatorname{sgn}}
\newcommand{\av}[1]{\left\langle #1\right\rangle}

\begin{document}

\title{Transport properties of semiconducting nanocrystal arrays at low temperatures}
\author{I.~S.~Beloborodov}
\affiliation{Materials Science Division, Argonne National
Laboratory, Argonne, Illinois 60439, USA}
\affiliation{Department of Physics, University of Chicago, Chicago, Illinois 60637, USA}

\author{A.~Glatz}\author{V.~M.~Vinokur}
\affiliation{Materials Science Division, Argonne National
Laboratory, Argonne, Illinois 60439, USA}
\date{\today}

\begin{abstract}
We study the electron transport in semiconducting nanocrystal arrays
at temperatures $T\ll E_c$, where $E_c$ is the charging energy for a
single grain. In this temperature range the electron transport is
dominated by co-tunneling processes. We discuss both elastic and
inelastic co-tunneling and show that for semiconducting nanocrystal
arrays the inelastic contribution is strongly suppressed at low
temperatures. We also compare our results with available
experimental data.
\end{abstract}

\maketitle

\newpage

Arrays of quantum semiconducting dots, artificial materials with
programmable electronic properties, can be tailored to fit many
applications ranging from solar cells and elements for energy
conversion to generation radiation detectors. Constructing
functional electronic devices based on nanocrystals requires
understanding their transport properties~\cite{Yu03,tran+prl05}. Electron
transport in nanocrystal arrays is governed by the interplay between
the internal energy structure of a single nanoparticle and by the
strength of inter-particle coupling, i.e by the probability of
electron tunneling between neighboring grains.  The coupling is
quantified by the \textit{tunneling
conductance}~\cite{BeloborodovRMP}, which can be tuned by varying
inter-granule distances in fabricated nanocrystal arrays (For review
see~\cite{Collier98,Murray00}).

Each semiconducting nanocrystal is characterized by two energy
scales: (i) the mean energy level spacing $\delta =1/(\nu a^d)$,
where $\nu$ is the density of states at the Fermi surface, $a$ is
the grain size, and $d$ is the dimensionality of a grain, and (ii)
the charging energy $E_c = e^2/\kappa a$ with $\kappa$ being the
effective dielectric constant (for a typical grain size of $a\approx
10\nm$ and $\kappa\approx 5$, $E_c$ are of the order of $300{\rm
K}$). In semiconductors the density of states $\nu$ is of about two
orders of magnitude smaller than that in metals. Thus in
semiconducting dots the mean energy level spacing, which is
inversely proportional to the density of states, can be of order of
the charging energy, $\delta \sim E_c$, in contrast to metallic
granular materials where typically $\delta\ll E_c$.

In this paper we consider transport properties of semiconducting
arrays in the limit of weak coupling between the grains, $g \ll
g_c$, where $g_c = 2e^2/h$ is the quantum conductance, and low
temperatures $T \ll \min \, (\delta, E_c)$. In this temperature
regime the electron transport due to sequential
tunneling~\cite{Averin89},
%\begin{equation}\label{eq.arrhenius}
following the activation law $\sigma \sim\exp(-E_c/T)$,
%\end{equation}
is strongly suppressed. Instead the conductivity of semiconducting
nanocrystal arrays is described by the following temperature
dependence~\cite{Philipe04,Wehrenberg05,Romero05,Yakimov03}:
\begin{equation}\label{eq.ES}
\sigma \sim \exp(-\sqrt{T_0/T}),
\end{equation}
where $T_0 = \alpha e^2/\kappa \xi$ is the characteristic energy
($T_0\gtrsim E_c$) scale with $\xi$ being the localization length
($\xi\lesssim a$) of the electron trapped on a particular
granule~\cite{remark_xi} and $\alpha$ a numerical constant of order
one. This behavior is identical to the temperature dependent {\it
Mott-Efros-Shklovskii} variable range hopping (VRH) conductivity in
usual bulk semiconductors~\cite{Shklovskiibook,Efros}.

\begin{figure}[h]
\includegraphics[width=0.8\linewidth]{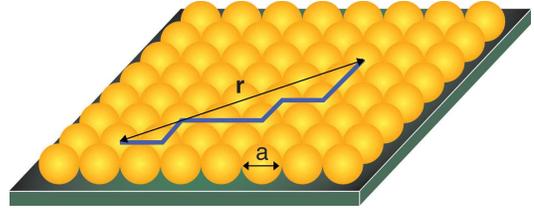}
\caption{Sketch of a two dimensional granular array with a schematic
multi-co-tunneling path through $N=r/a$ grains with $a$ and $r$
being the grain size and the hopping length,
respectively.}\label{fig.grains}
\end{figure}

Two remarks are in order: First, this VRH behavior can be observed
at temperatures $T < T_0$. At temperatures $T \sim T_0$ the hopping
length $r$ becomes of the order of the grain size $a$, see
Fig.~\ref{fig.grains}, and Eq.~(\ref{eq.ES}) does not hold anymore.
Close to $T_0$, the conductivity follows the simple activation law.
Second, Eq.~(\ref{eq.ES}) for the conductivity holds also for three
dimensional semiconducting nanocrystal arrays and thick granular
films with the sample thickness $L$ larger than the hopping length
$r$.

For the  VRH process of Eq.~({\ref{eq.ES}}) to realize,  two
ingredients are necessary: first, a finite density of localized
states (DOS) on the Fermi surface and second, the mechanism for an
electron to tunnel over a distance exceeding the granule size, i.e.
a mechanism for a ``direct" tunneling through many grains (see
Fig.~\ref{fig.grains}). The possible source of a finite DOS in a
periodic granular array are charged impurities in the insulating
matrix which generate random shifts of the chemical potential of a
granule; this was suggested in Ref.~[\onlinecite{Shklovskii04}]. The
tunneling mechanism mediating VRH was identified as the so-called
multiple co-tunneling process in
Refs.~[\onlinecite{Beloborodov05,Feigelman05}], where a theory for
VRH in arrays of metallic granules was constructed.

\begin{figure}[h]
\includegraphics[width=0.9\linewidth]{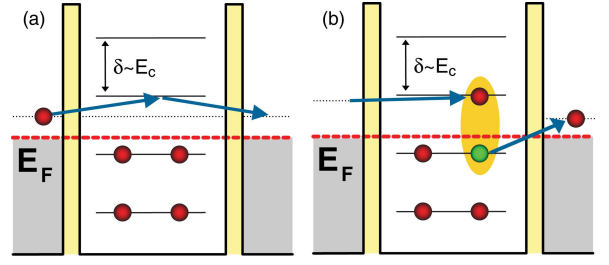}
\caption{(a) elastic co-tunneling process. In this process the
charge is transferred via the tunneling of an electron through an
intermediate virtual state in the dot such that the electron leaves
the dot with the same energy as it came. (b) inelastic co-tunneling
process. In this process the electron coming out of the dot has a
different energy than the incoming one. During this process an
electron-hole excitation is left behind in the grain, which absorbs
the energy difference of the in- and out-going
electron.}\label{fig.cotunnel}
\end{figure}

The essence of a co-tunneling process is that an electron tunnels
via virtual states in intermediate granules thus bypassing the huge
Coulomb barrier.  This can be visualized as coherent superposition
of two events: tunneling of the electron into a granule and the
simultaneous escape of another electron from the same granule. There
are two distinct mechanisms of co-tunneling processes,
\textit{elastic} and \textit{inelastic}
co-tunneling~\cite{Averin90,Glazman06}. Elastic co-tunneling means
that the electron that leaves the dot has the same energy as the
incoming one, Fig.~\ref{fig.cotunnel}~(a). In the event of inelastic
co-tunneling, the electron coming out of the dot has a different
energy than the entering electron. This energy difference is
absorbed by an electron-hole excitation in the granule, which is
left behind in the course of the inelastic co-tunneling process,
Fig.~\ref{fig.cotunnel}~(b).

The probability for electron tunneling through many grains via
elastic or inelastic co-tunneling can be most easily found for the
case of the diagonal Coulomb interaction, namely
\begin{equation}\label{eq.probability}
P  = \prod\limits_{i=1}^{N} P_i,
\end{equation}
where $P_i$ is the probability of one elastic/inelastic co-tunneling
event through a single grain and $N=r/a$ is the number of grains,
Fig.~\ref{fig.grains}. Below we discuss  the probability $P_i$ for
elastic and inelastic electron co-tunneling separately and derive
expressions for the corresponding localization length $\xi$, which
is defined by $P\sim e^{-r/\xi}$. In terms of the amplitude $A_i$
for an electron tunneling through a grain the probability can be
expressed as $P_i \sim \av{ |A_i|^2 }$, where the symbol $\av{...}$
denotes the averaging over randomness in the system.

Semiconducting nanocrystal arrays are described by the Hamiltonian
\begin{equation}
{\cal H}=\sum\limits_i {\cal H}^{(i)}+\sum_{\av{ij}}{\cal
H}_T^{(ij)}\,, \label{eq.ham}
\end{equation}
where $i,j$ are the grain indexes and summation in the second term
of the r.h.s. of Eq.~(\ref{eq.ham}) is performed over nearest
neighbors. The term ${\cal H}^{(i)}$ stands for electrons in the
single grain $i$,  and ${\cal H}_T^{(ij)}$ is the tunneling
Hamiltonian between the adjacent grains $i$ and $j$
\begin{equation}
{\cal H}_T^{(ij)}={\sum\limits_{k_i,k_j}}^{\prime} t_{ij}{\hat
c}_{k_i}^{(i)\dagger}{\hat c}_{k_j}^{(j)}+\textrm{h.c.}\,.
\label{eq.tham}
\end{equation}
Here $t_{ij}$ are the random tunneling matrix elements and ${\hat
c}_{k_i}^{(i)\dagger}$ [${\hat c}_{k_i}^{(i)}$] is the creation
[annihilation] operator on the $i$th grain. The symbol ${}^{\prime}$
at the sum is to reflect that due to the large mean level spacing in
semiconducting grains $\delta \sim E_c$, only a few terms of the sum
are important.

{\it Elastic co-tunneling.---} First we discuss the conductivity of
semiconducting nanocrystal arrays due to elastic co-tunneling,
Fig.~\ref{fig.cotunnel}~(a). In order to derived the probability
$P_i^{\el}$ for such a process, we calculate the related amplitude
for the elemental elastic co-tunneling process which is in general
given by $A^{\el} = \nu {\sum\limits_k}^{\prime} |t|^2 \sgn(
\varepsilon_k)/(E_c + |\varepsilon_k|)$; the energies
$\varepsilon_k$ are differences of energy levels of adjacent grains.
Averaging over the random tunneling matrix elements, we obtain for
the amplitude probability density the following expression
\begin{equation}\label{eq.Ael2}
\av{|A^{\el}|^2}  = {\sum\limits_k}^{\prime} \frac{1}{\tau^2(E_c +
|\varepsilon_k|)^2 },
\end{equation}
where $\tau^{-1} = \nu \, \av{ |t|^2}$ is the inverse electron
escape time from a grain and we used Gaussian distributed tunneling
matrix elements~\cite{remark_distr},
$\av{t_{ij}t_{kl}^{\ast}}=\av{|t|^2}(\delta_{ik}\delta_{jl} +
\delta_{il}\delta_{jk})$.  Since for semiconducting grains the mean
level spacing $\delta$ is of the order of
$E_c$~[\onlinecite{remark_sum}], we obtain for the elastic
localization length $\xi^{el}$, using Eqs.~(\ref{eq.probability}),
(\ref{eq.Ael2}), the following expression
\begin{equation}\label{eq.xiel}
\xi^{el} \sim \frac{a}{\ln [ \tau \, \max (E_c, \delta) ]^2}\,.
\end{equation}
For $\delta\ll E_c$ one has to take into account all terms of the
sum in (\ref{eq.Ael2})  leading to a different power of the
logarithm in the denominator of (\ref{eq.xiel}).
 Therefore it follows that for $\delta_{\text{metallic}}\ll
E_c\lesssim \delta_{\text{semiconducting}}$  the localization length
of semiconducting nanocrystal arrays is smaller than the
localization length in metallic arrays if the grain size and
dielectric constant, i.e. $E_c$, are the
same~\cite{Beloborodov05,BeloborodovRMP}.

{\it Inelastic co-tunneling.---}Next, we discuss the contribution to
the conductivity of a semiconducting nanocrystal array due to
inelastic co-tunneling processes, Fig.~\ref{fig.cotunnel}~(b). In
contrast to the elastic process the amplitude has an exponential
dependence on the mean level spacing, $A^{\inel} \sim (\nu \,
|t|^2/E_c) \, \exp(-\delta/T)$, which reflects the fact that the
smallest energy of an electron-hole pair is of the order of mean
energy level spacing in a grain. The amplitude probability density
for an inelastic co-tunneling process has therefore the form
\begin{equation}\label{eq.Ainel}
\av{|A^{\inel}|^2} \sim \frac{1}{\tau^2}\,
\frac{\exp(-2\delta/T)}{[\max(\delta, E_c)]^2} \, .
\end{equation}
For high temperatures, $T \gg \delta$, the numerator approaches
unity. For the inelastic localization length $\xi^{\inel}$ we obtain
\begin{equation}\label{eq.xiinel}
 \xi^{\inel} \sim \frac{a}{\ln[\, \tau \, {\rm max} (\delta, E_c)\,]^2 + 2\delta/T}\,.
\end{equation}
For temperatures $T < \delta/\ln[\, \tau \, {\rm max} (\delta,
E_c)\,]^2$, the last term in the denominator of
Eq.~(\ref{eq.xiinel}) is the dominant one. In this case the
localization length reduces to
\begin{equation}\label{eq.xiinellow}
\xi^{\inel} \sim a\,  (T/\delta).
\end{equation}
Comparing Eqs.~(\ref{eq.xiel}) and (\ref{eq.xiinellow}) one can see
that $\xi^{\inel} \ll \xi^{\el}$ at low temperatures, meaning that
inelastic processes are strongly suppressed and therefore the main
mechanism for conductivity is elastic co-tunneling.

{\it Discussion.---}As we have shown, the inelastic co-tunneling
process is suppressed in semiconducting nanocrystal arrays. This can
also be seen by comparing the crossover temperatures separating the
elastic and inelastic mechanisms of electron co-tunneling in the
metallic and semiconducting case. For metallic granular arrays this
is of the order of $\sqrt{E_c\,
\delta}$~[\onlinecite{Averin90,BeloborodovRMP,Feigelman05}], where
the mean level spacing $\delta$ is small. In the case of
semiconducting nanocrystal arrays $\delta$ is of the order of the
charging energy, $\delta \sim E_c$, therefore the crossover
temperature is of the order of $E_c$, i.e., much larger than the
temperature $T$ such that at low temperatures inelastic co-tunneling
does not contribute to the transport.

If the applied electric field grows sufficiently high, $E
> T/ e \xi$, the hopping conductivity becomes field-dependent:
$\sigma \sim \exp(-\sqrt{E_0/E})$, where $E_0 =\alpha e/\kappa\xi^2$
and $E_0>E$, in full analogy with the Shklovskii
result~\cite{Shklovskii73} for hopping conductivity in usual bulk
semiconductors in a strong electric field.

As a last remark, we would like to discuss the possibility of the
observation of the Mott law in semiconducting nanocrystal arrays. In
usual (bulk) semiconductors, the Efros-Shklovskii law may turn into
the Mott behavior with the increase of temperature. This happens
when the typical electron energy involved in a hopping process
becomes larger than the width of the Coulomb gap $\Delta_c$, i.e.
when it falls into the flat region of the density of states where
Mott behavior is expected. To estimate the width of the Coulomb gap,
$\Delta_c$, one compares the Efros-Shklovskii expression for the
density of states
\begin{equation}
\nu(\Delta_c) \sim (\kappa/e^2)^d |\Delta_c|^{d-1}
\end{equation}
with the DOS in the absences of the long-range part of the Coulomb
interactions, $\nu_0$. Using the condition $\nu(\Delta_c) \sim
\nu_0$ we obtain
\begin{equation}\label{eq.crossover}
 \Delta_c = \left( \frac{\nu_0 e^{2d}}{\kappa^d} \right)^{1/(d-1)} .
\end{equation}
Inserting the value for the bare DOS, $\nu_0 = [\max (\delta, E_c)
\, a^d]^{-1}$, into Eq.~(\ref{eq.crossover}) we finally obtain
\begin{equation}\label{eq.delta}
 \Delta_c \sim E_c\left[\frac{E_c}{\max(E_c,\delta)}\right]^{\frac{1}{d-1}}\,.
\end{equation}
Equation~(\ref{eq.delta}) means that there is no flat region in the
density of ground states and, thus, the Mott regime is difficult to
observe in semiconducting nanocrystal arrays.

Recent experiments on semiconducting nanocrystal materials, CdSe,
PbSe, and GeSi, revealed the VRH
conductivity~\cite{Philipe04,Wehrenberg05,Romero05,Yakimov03}. The
results allow for determining the characteristic energy scale $T_0$
and/or field $E_0$. These scales are related via $T_0=e E_0\xi$.
Using our results for the localization length (\ref{eq.xiel}) in
connection with the definition of both energy scales, e.g., the mean
grain size $a$ can be calculated, i.e., represents an alternative to
direct measurements with, e.g. a scanning tunneling microscope. As
an example we reexamine the results of Ref.~[\onlinecite{Philipe04}]
for CdSe nanocrystal arrays which were measured in the low and high
electric field regime: $T_0$ was measured to be $5200{\rm K}$ which
gives together with $\kappa\sim 4$ a localization length $\xi\sim
1\nm$. Using their value for $E_0\sim 7.5\cdot 10^7{\rm V}/{\rm m}$
gives $\xi\sim 2\nm$. Due to the logarithm in (\ref{eq.xiel}), the
localization length is of the order of the grain size ($\xi\leq a$)
and therefore our estimates are also in agreement with their
directly observed size of $5.4\nm$.

In conclusion, we have discussed transport properties of
semiconducting nanocrystal arrays at low temperatures. We have shown
that the electron transport is dominated by co-tunneling processes
and for temperatures $T \ll {\rm min} (\delta, E_c)$ the main
mechanism for electron tunneling through many grains is the elastic
electron co-tunneling. Our results for the localization length $\xi$
can be used to extract information about the morphology of the
sample, e.g., the average grain size, without direct measurements.

{\it Acknowledgements.---}We thank Philippe Guyot-Sionnest and
Andrei Lopatin for useful discussions. A.G. acknowledges support by
the DFG through a research grant. This work was partly supported by the U.S. Department of Energy Office
of Science through contract No. DE-AC02-06CH11357.


\begin{thebibliography}{99}

\bibitem{Yu03} D.~Yu, C.~Wang, and P.~Guyot-Sionnest, Science {\bf
300}, 1277 (2003).

\bibitem{tran+prl05} T. B. Tran, I. S. Beloborodov, X. M. Lin, T. P. Bigioni, V. M. Vinokur,
and H. M. Jaeger, Phys. Rev. Lett. {\bf 95}, 076806 (2005).

\bibitem{BeloborodovRMP} I.~S.~Beloborodov, K.~B.~Efetov, A.~Lopatin, and V.~M.~Vinokur,
cond-mat/0603522 (to be published in Rev. Mod. Phys.).

\bibitem{Collier98} C.~Collier, T.~Vossmeyer, and J.~Heath, Annual review of
Physical Chemistry {\bf 49}, 371 (1998).

\bibitem{Murray00} C.~Murray, C.~Kagan, and M.~Bawendi, Annual review of
Materials Science {\bf 30}, 545 (2000).

\bibitem{Averin89} D.~V.~Averin and A.~A.~Odintsov, Phys. Lett. A~\textbf{140}, 251 (1989).


\bibitem{Philipe04} D.~Yu,  C.~J.~Wang, B.~L.~Wehrenberg, and P.~Guyot-
Sionnest, Phys. Rev. Lett.~\textbf{92}, 216802 (2004).

\bibitem{Wehrenberg05} B.~L.~Wehrenberg, D.~Yu, J.~Ma, and
P.~Guyot-Sionnest, J. Phys. Chem. B~{\bf 109}, 20192 (2005).

\bibitem{Romero05} H.~Romero and M.~Drndic, Phys. Rev. Lett.~\textbf{95},
156801 (2005).

\bibitem{Yakimov03} A. I. Yakimov, A. V. Dvurechenskii, A. V. Nenashev, and A. I.
Nikiforov, Phys. Rev. B {\bf 68}, 205310 (2003).

\bibitem{remark_xi} On length scales well
exceeding the grain size (for example on the scale of the hopping
length) the behavior of the semiconducting nanocrystal array is
similar to that of bulk semiconductors.

\bibitem{Shklovskiibook} B.~I.~Shklovskii and A.~L.~Efros, {\it Electronic properties
of Doped Semiconductors, Springer-Verlag, New York, 1988}.

\bibitem{Efros} A.~L.~Efros and B.~I.~Shklovskii, J.~Phys.~C \textbf{8}, L49
(1975).

\bibitem{Shklovskii04} J.~Zhang and B.~I.~Shklovskii,
Phys. Rev. B~{\textbf  70}, 115317 (2004).

\bibitem{Feigelman05} M.~V.~Feigel'man and A.~S.~Ioselevich,
Pis'ma Zh. Eksp. Teor. Fiz. {\bf 81}, 341 (2005) [Sov. Phys. JETP
Lett. {\bf 81}, 227 (2005)].

\bibitem{Beloborodov05} I.~S.~Beloborodov, A.~Lopatin, and V.~Vinokur,
Phys. Rev. B~\textbf{72}, 125121 (2005).

\bibitem{Averin90} D.~Averin and Yu.~Nazarov, Phys. Rev. Lett. \textbf{65}, 2446 (1990).

\bibitem{Glazman06} L.~Glazman and M.~Pustilnik. cond-mat/0501007

\bibitem{remark_distr} For our considerations the assumption of a Gaussian distribution of
the tunneling matrix elements is not crucial, since a different
distribution will only change the expression for the electron escape
time $\tau$ and therefore the qualitative picture for the
localization length $\xi$ remains the same.

\bibitem{remark_sum} Eq.~(\ref{eq.xiel}) is valid also if $\delta$
is slightly smaller than $E_c$. Only in this case one can take into
account only a few terms in the summation over the states $k$ in
Eq.~(\ref{eq.Ael2}) in order to derive Eq.~(\ref{eq.xiel}). If the
grains are metallic Eq.~(\ref{eq.xiel}) is still valid as long as
the grain size is very small such that $\delta$ is of order $E_c$.

\bibitem{Shklovskii73} B.~I.~Shklovskii, Fiz. Tekh. Poluprovodn.
(S.-Petersburg) {\bf 6}, 2335 (1973) [Sov. Phys. Semicond. {\bf
6},1964 (1973)].

\end{thebibliography}
\end{document}